\documentstyle[art12,psfig,epsf]{article}
\begin{document}
\baselineskip 21pt

\begin{center}
 {\large\bf Spin effects in two-particle hadronic\\
            decays of $B_c$ ~mesons.}

\vspace{4mm}

{\sl Pakhomova O.N., Saleev V.A.}\footnote
{Email: saleev@ssu.samara.ru}\\[2mm]
Samara State University, 443011, Samara\\
and\\
Samara Municipal Nayanova
University, 443001, Samara\\
Russia
\end{center}

\begin{abstract}
We consider spin effects in two--particle hadronic decays of
$B_c$ and $B_c^*$ ~mesons into $J/\psi$ plus $\rho(\pi)$
in the frame work of hard gluon exchange model.
It is shown that polarization of the $J/\psi$ ~meson
is very different in decays of
$B_c$ and $B_c^*$ ~mesons as well as their decay widths into
$J/\psi$ plus $\rho(\pi)$.
\end{abstract}

\subsection*{1. Introduction}
The start of the experimental study of the $B_c$ ~mesons, containing heavy
quarks of different flavors, was done recently by CDF Collaboration at
FNAL ~\cite{1}. The obtained value of
$B_c$ ~meson mass is $m_{B_c}=6.40\pm 0.39\pm 0.13$ GeV.
It doesn't contradict theoretical predictions, which have been made in
nonrelativistic potential models ~\cite{2} and QCD sum rules method
~\cite{3}.
However, the predicted value for the difference between masses
$^3S_1$ state ($B_c^*$) and $^1S_0$ state ($B_c$) of the ($\bar bc$)-- system
is $0.05-0.07$ GeV, that now much  smaller than experimental uncertainties
~\cite{1}.

The calculation, which was done in $\alpha_s^4$
order of perturbative QCD using nonrelativistic model for heavy quarkonium,
shows that the ratio of
$B_c^*$ and $B_c$ ~mesons production cross sections in high energy hadronic
collisions is about $2\sim 3$ ~\cite{4}.
At the same time, the fragmentation model of heavy quarkonium production
predicts that $\sigma(B_c^*)/\sigma(B_c)
\approx 1.3$ ~\cite{5}.
Such a way, experimental separation of
$B_c^*$ and $B_c$~meson production cross section is very important for test
of the models, which pretend on description of  $B_c$ ~meson production.

The signal of $B_c$ ~meson production in ~\cite{1}
is a peak in the invariant mass spectrum of
$J/\psi l \nu_l (l=e,\mu)$,
system, which was produced in semileptonic decay
\begin{equation}
B_c\to J/\psi+l+\nu_l, \label{1}
\end{equation}
where $J/\psi$ was detected using lepton decay mode
$(J/\psi\to l^+l^-).$
However, more suitable decay channels from the registration point of view,
are two-particle hadronic decays with
$J/\psi$~meson in final state:
\begin{equation}
B_c(B_c^*)\to J/\psi\rho(\pi),\label{2}
\end{equation}
It is assumed, that branching ratio of the decays (\ref{2})
is about 1\% ~\cite{6}.

The main feature of the two-particle hadronic $B_c$ decays is  the fact
that both quarks in the $B_c$ ~meson
($\bar b$ and spectator $á$--quark) are heavy.
The necessary condition of
$c\bar c$ bound state formation creation is large momentum transfer
$(k^2>>\Lambda_{QCD}^2)$  to the spectator quark
and spectator picture is not valid in the decays under consideration.
In fact, the gluon virtuality in the decays (\ref{2})
is equal to
$$k^2\approx-\frac{m_2}{4m_1}((m_1-m_2)^2-m_3^2)\approx
-1.2~\mbox{GeV}^2,$$
where $m_1$ is $B_c$ or $B_c^*$ ~meson mass, $m_2$ is  $J/\psi$ ~meson
mass and $m_3$ is $\rho$ or $\pi$ ~meson mass. So,
it is needed to use here the hard scattering formalism ~\cite{7}.

\subsection*{2. The hard exchange model}
At first time, the exclusive hadronic decays
$B_c^+\to J/\psi\pi^+(\rho^+)$ or $B_c^+\to\eta_c\pi^+(\rho^+)$
have been under consideration in the frame work of hard exchange model in
paper ~\cite{6}.
It was shown, that in contrast to the spectator approach, the hard
$t$--channel exchange results in the approximate double enhancement of the
decay amplitudes. However the trivial arithmetic error in
$\Gamma(B_c^+\to \eta_c\pi^+)$ calculation was made in~\cite{6}.
The decay widths of $B_c^*$ ~mesons,
as well as spin asymmetry for the creation of $J/\psi$ ~mesons in the
states with different polarization didn't were calculated.
The polarization of the $J/\psi$ ~meson in semileptonic decay of $B_c$
~meson ($B_c^+\to J/\psi+\mu+\bar\nu_{\mu}$)
was studied recently in \cite{7b} using the spectator approach.

In the frame work of hard gluon exchange model, the amplitudes for decays
$B_c(B_c^*)\to J/\psi\rho(\pi)$ are described by diagrams in Fig.1.
Using nonrelativistic approach, we assume that quarks in a heavy quarkonium
move with equal velocity.
So, in the case of $B_c$ ~meson with four--momentum $p_1$ one has
$$p_c=\frac{m_c}{m_b+m_c}p_1,\qquad p_{\bar b}=\frac{m_b}{m_b+m_c}p_1,$$
and in the case of $J/\psi$ ~meson with four--momentum $p_2$
$$p_c=\frac{1}{2}p_2,\qquad p_{\bar c}=\frac{1}{2}p_2.$$
Assuming that quarkonium binding energies are small,
the heavy quark masses may
be presented in the terms of  $m_1$ and $m_2$ as follows
$$m_c=\frac{m_2}{2},\qquad m_b=m_1-\frac{m_2}{2}.$$
The transition from the amplitude of free heavy quarks with equal
four--velocity $v=p/m$
production (or decay) to the amplitude of the binding state production
(or decay) is done using following prescription:
$$V^i(v)\bar U^j(v)\to \hat a\frac{(1+\hat v)}
{2}\frac{f m}{2\sqrt 3}
\frac{\delta^{ij}}{\sqrt 3},$$
where $\hat a=\gamma_{\mu}\varepsilon^{\mu}$,
$\varepsilon^{\mu}$ is four--vector of $B_c^*$ ~meson polarization,
or $\hat a=\gamma^5$ in the case of psevdoscalar quarkonium,
$f$ is the leptonic constant, which is related with the value of
configuration wave function at the origin
$$f=\sqrt{\frac{12}{m}}\vert \psi(0)\vert,$$
$\delta^{ij}/\sqrt 3$ is the color factor, which takes into account color
singlet state of heavy quarks in quarkonium.

Because of the quark virtuality in the decays (\ref{2})
is very small in compare with $W$--boson mass, one can use four-fermion
effective theory and write $\bar b\to\bar c\pi^+(\rho^+)$ vertex in the
terms of effective Fermi--constant $G_F/\sqrt 2$, mixing parameter
$V_{bc}$, left quark current and axial--vector current of
$\pi$~meson ($f_3 p_{3}^\mu, f_{3}=f_{\pi})$ or $\rho$~meson
($m_3f_3\varepsilon_{3}^\mu, f_{3}=f_{\rho})$.

For example, the amplitude of $B_c^*\to J/\psi\rho$ decay calculated
according to the diagrams in
Fig.1, take the form
\begin{eqnarray}
&&M(B_c^*\to J/\psi\rho)=\frac{G_F}{\sqrt 2}V_{bc}4\pi\alpha_s
\frac{4}{3}m_3f_3 \frac{f_1m_1}{2\sqrt 3} \frac{f_2m_2}{2\sqrt 3}
\frac{a_1}{8k^4}\times\nonumber\\
&&Tr\biggl [\hat\varepsilon_2(1+\hat v_2)\gamma^\alpha
\hat\varepsilon_1(1-\hat v_1)
\biggl (\hat\varepsilon_3(1-\gamma_5)(-\hat x_1+m_c)\gamma^\alpha+\\
\label{3}
&&\frac{m_2}{m_1}\gamma^\alpha(-\hat x_2+m_b)\hat\varepsilon_3(1-\gamma_5)
\biggr )\biggr ],     \nonumber
\end{eqnarray}
where
\begin{eqnarray}
&&\hat x_1=m_2\hat v_2-\frac{m_2}{2}\hat v_1,\\ \nonumber
&&\hat x_2=m_1\hat v_1-\frac{m_2}{2}\hat v_2,\\ \nonumber
&&k^2=\frac{m_2^2}{2}(1-y),\\ \nonumber\label{4}
&&y=(v_1v_2)=\frac{m_1^2+m_2^2-m_3^2}{2m_1m_2}. \nonumber
\end{eqnarray}
The factor $a_1$ comes from hard gluon corrections to the four-fermion
effective vertex.
In the case of psevdoscalar $B_c$~meson decays it is needed to do
substitution
$\hat\varepsilon_1\to
\gamma_5$, and, if one has $\pi$~meson in the final state,
$\hat\varepsilon_3\to
\hat v_3$.
There are following cinematic formulae in the rest frame of $B_c$~meson:
\begin{eqnarray}
&&v_1=(1,0,0,0),\\ \nonumber
&&v_2=\frac{1}{m_2}(E_2,0,0,\vert\vec{p_2}\vert),\\ \nonumber
&&v_3=\frac{1}{m_3}(m_1v_1-m_2v_2),\\ \nonumber
&&E_2=\frac{m_1^2+m_2^2-m_3^2}{2m_1},\\ \nonumber\label{5}
&&\vert\vec{p_2}\vert=\sqrt{E_2^2-m_2^2}. \nonumber
\end{eqnarray}
Put $m_3=0$ in (\ref{5}), one finds:
$$
E_2\approx\frac{m_1^2+m_2^2}{2m_1}\qquad{\rm ¨}\qquad
\vert \vec p_2\vert \approx\frac{m_1^2-m_2^2}{2m_1}.
$$
The summation over polarization states of vector particles may be done using
formula
\begin{equation}
\sum_{ L,T}\varepsilon_\mu^*(v)\varepsilon_\nu(v)
=-g_{\mu\nu}+v_\mu v_\nu.
\end{equation}
The longitudinal polarization
four--vector $\varepsilon_{ L}(p)$ is usually written explicitly as
\begin{equation}
\varepsilon_{ L}^\mu(p)=\left(\frac{\vert\vec p\vert}{m},
\frac{E\vec p}{m\vert\vec p\vert}\right)
\end{equation}
Let us definite an auxiliary four--vector
$n^\mu=(1,-\vec p/\vert\vec p\vert)$ such that $n^2=0,\ (np)=E+\vert\vec
p\vert$. With the help of four--vector $n^\mu$ one can rewrite
$\varepsilon_{ L}^\mu(p)$ in the following covariant form
\begin{equation}
\varepsilon_{ L}^\mu(p)=\frac{p^\mu}{m}-\frac{mn^\mu}{(np)},
\end{equation}

We can then obtain following covariant expressions for the longitudinal and
transverse polarization sum:
\begin{eqnarray}
&&\varepsilon_{ L}^{*\mu}(v)\varepsilon_{ L}^{\nu}(v)
=v^\mu v^\nu-\frac{v^\mu n^\nu+v^\nu n^\mu}{(vn)}
+\frac{n^\mu n^\nu}{(vn)^2},\\
&&\sum_{ T}\varepsilon_{ T}^{*\mu}(v)
\varepsilon_{ T}^{\nu}(v)
=-g_{\mu\nu}+\frac{v^\mu n^\nu+v^\nu n^\mu}{(vn)}
-\frac{n^\mu n^\nu}{(vn)^2}.
\end{eqnarray}

\subsection*{3. The results}
The unpolarized width of $B_c^*\to J/\psi\rho$ decay is written in the
following form
\begin{equation}
\Gamma(B_c^*\to J/\psi\rho)=\frac{\vert\vec p_2\vert}
{8\pi m_1^2}
\overline{\vert M\vert^2}.
\end{equation}
where $\overline{\vert M\vert^2}$ is squared matrix element after summation
over polarizations of $J/\psi$, $\rho^+$ and $B_c^*$ ~mesons, divided by 3
(the number of the initial $B_c^*$~meson polarization states).
From Eqs. (3) -- (11) one gets:
\begin{equation}
\Gamma(B_c^*\to J/\psi\rho)=G_F^2|V_{bc}|^2\frac{4\pi\alpha_s^2}
{81}\frac{f_1^2f_2^2f_3^2}{m_1^4m_2^6}\frac{|\vec{p_2}|}{(1-y)^4}
a_1^2F(B_c^*),
\end{equation}
\begin{equation}
\Gamma(B_c\to J/\psi\rho)=G_F^2|V_{bc}|^2\frac{4\pi\alpha_s^2}
{81}\frac{f_1^2f_2^2f_3^2}{m_1^4m_2^6}\frac{|\vec{p_2}|}{(1-y)^4}
a_1^2F(B_c),
\end{equation}
where $f_1=f_{B_c}=f_{B_c^*}$, $f_2=f_{J/\psi}$, $f_3=f_\rho$,
$m_1=m_{B_c}$, $m_2=m_{J/\psi}$, $m_3=m_{\rho}$ and
\begin{eqnarray*}
&&3\times F(B_c^*)=m_3^8+2m_3^6(13m_1^2-2m_1m_2-m_2^2)\\
&&+2m_3^4(-19m_1^4-18m_1^3m_2-16m_1^2m_2^2+2m_1m_2^3+m_2^4)\\
&&+2m_3^2(-3m_1^6+18m_1^5m_2+m_1^4m_2^2-
20m_1^3m_2^3+11m_1^2m_2^4+2m_1m_2^5-m_2^6)\\
&&+17m_1^8+4m_1^7m_2-32m_1^6m_2^2-
12m_1^5m_2^3+14m_1^4m_2^4+\\
&&+12m_1^3m_2^5-4m_1m_2^7+m_2^8,
\end{eqnarray*}
\begin{eqnarray*}
&&F(B_c)=m_3^8+2m_3^6m_1(5m_1+8m_2)\\
&&+m_3^4(-19m_1^4-56m_1^3m_2+6m_1^2m_2^2-8m_1m_2^3-3m_2^4)\\
&&+2m_3^2(2m_1^6+20m_1^5m_2-3m_1^4m_2^2-
16m_1^3m_2^3+16m_1^2m_2^4-4m_1m_2^5+m_2^6)\\
&&+4m_1^4(m_1^4-2m_1^2m_2^2+m_2^4).
\end{eqnarray*}
In the limit of vanishing $\rho$ ~meson mass
and $m_{B_c^*}=m_{B_c}$, one gets:
\begin{eqnarray}
&&\frac{\Gamma(B_c^*\to
J/\psi\rho)}{\Gamma(B_c^*\to J/\psi\pi)}\approx
\frac{\Gamma(B_c\to
J/\psi\rho)}{\Gamma(B_c\to J/\psi\pi)}\approx
\biggl (\frac{f_{\rho}}{f_{\pi}}\biggr )^2=2.47\\[3mm]
&&\frac{\Gamma(B_c^*\to J/\psi\rho)}
{\Gamma(B_c\to J/\psi\rho)}\approx\
\frac{\Gamma(B_c^*\to J/\psi\pi)}
{\Gamma(B_c\to J/\psi\pi)}
=\frac{1}{12}
(17+4x+2x^2-4x^3+x^4)\approx 1.58\nonumber,
\end{eqnarray}
where $x=m_2/m_1$, $m_1=6.3$ GeV, $m_2=3.1$ GeV.

The explicit calculation with $m_3=m_{\rho}= 0.77$ GeV, $m_{B_c^*}
= 6.3$ GeV and $m_{B_c}= 6.25$ GeV, gives following results:
\begin{equation}
\frac{\Gamma(B_c^*\to J/\psi\rho)}
{\Gamma(B_c\to J/\psi\rho)}\approx 1.43,
\end{equation}
and
\begin{equation}
\frac{\Gamma(B_c^*\to
J/\psi\rho)}{\Gamma(B_c^*\to J/\psi\pi)}\approx 2.28, \qquad
\frac{\Gamma(B_c\to
J/\psi\rho)}{\Gamma(B_c\to J/\psi\pi)}\approx 2.54
\end{equation}
The polarization of $J/\psi$ ~meson in the decays (\ref{2}) can be measured
using angular distribution of the leptons in the decay
$J/\psi\to l^+l^-$. In the rest frame of
$J/\psi$ ~meson the angular distribution of the leptons is given by
\begin{equation}
\frac{d\Gamma}{d\theta}(J/\psi\to l^+l^-)\sim 1+\alpha\cos^2\theta,
\end{equation}
where $\theta$ is the angle between the outgoing lepton 3--momentum and the
polarization axis of $J/\psi$ ~meson and
$$\alpha=\frac{\Gamma_T-2\Gamma_L}{\Gamma_T+2\Gamma_L},$$
$\Gamma_T$ and $\Gamma_L$ are decay widths of $B_c^*$ ~meson into the
transverse and longitudinal polarized $J/\psi$ ~meson.
The exact results, taking into account $\rho$ ~meson mass, are following
\begin{eqnarray}
&&\alpha(B_c^*\to J/\psi\rho)\approx 0.40\\
&&\alpha(B_c\to J/\psi\rho)\approx-0.85
\end{eqnarray}
In the limit of $m_\rho=0$, one gets very simple expressions
\begin{equation}
\alpha(B_c^*\to J/\psi\rho)\approx\alpha(B_c^*\to
J/\psi\pi)\approx\frac{7-4x-2x^2+4x^3-x^4}{9+4x+2x^2-4x^3+x^4}= 0.45
\end{equation}
and
\begin{equation}
\alpha(B_c\to J/\psi\rho)\approx\alpha(B_c\to
J/\psi\pi)=-1.
\end{equation}

The probability of the spin conservation  during $B_c^*$
~meson decay into $J/\psi$ ~meson is described by the parameter
\begin{equation}
\xi=\frac{\Gamma_{T\rightarrow T}}{\Gamma_{T\rightarrow T}+
\Gamma_{T\rightarrow L}},
\end{equation}
where $\Gamma_{T\rightarrow T}$ is the decay width of the
transverse polarized
$B_c^*$ ~meson into transverse polarized $J/\psi$ ~meson and
$\Gamma_{T\rightarrow L}$ is the decay width of the transverse polarized
$B_c^*$~meson into longitudinal polarized
$J/\psi$~meson. It is obviously that in the limit of vanishing
$\rho$ ~meson mass, one has
\begin{equation}
\xi(B_c^*\to J/\psi\rho)\approx\xi(
B_c^*\to J/\psi\pi)=1.
\end{equation}
Taking into account $\rho$ ~meson mass, one finds
$$\xi(B_c^*\to J/\psi\rho)=0.97$$

Using following set of parameters ~\cite{8}:
$f_{B_c}=f_{B_c^*}=0.44\mbox {~GeV},\ f_{J/\psi}=0.54\mbox {~GeV}, \
f_\pi=0.14\mbox {~GeV}$, $f_\rho=0.22\mbox {~GeV},\ V_{bc}=0.04$, $
G_F=1.166\cdot10^{-5}\mbox {~GeV}^{-2}$ and $\alpha_s\approx 0.33,$
we have obtained  (in units $10^{-6}$ eV)
\begin{eqnarray}
&& \Gamma (B_c\to J/\psi\rho)\approx 22.3 a_1^2,\\
&& \Gamma (B_c^*\to J/\psi\rho)\approx 31.9 a_1^2,\\
&& \Gamma (B_c\to J/\psi\pi)\approx 8.76 a_1^2 ,\\
&& \Gamma (B_c^*\to J/\psi\pi)\approx 13.9 a_1^2.
\end{eqnarray}
Our results
(24) -- (27) agree with the values obtained in ~\cite{8,9} for
$B_c\to J/\psi\rho(\pi)$ decays, if we take into consideration
additional factor
$a_1^2\approx1.21$, which comes from the hard QCD corrections
to the  effective four--fermion vertex.
Note, that the decay width for $B_c^*\to J/\psi \rho(\pi)$ is much
smaller than the gamma decay width for $B_c^*$ meson,
which is about 60 eV \cite{8}. This one makes impossible experimental
study the weak decays of $B_c^*$ mesons.

\subsection*{4. Conclusion}
We have shown that the widths of vector
$B_c^*$ ~meson decays $\Gamma(B_c^*\to J/\psi\rho)$ and
$\Gamma(B_c^*\to J/\psi \pi)$, in 1.43 and 1.67 times larger then
corresponding decay widths of scalar $B_c$ ~meson.

In decays of $B_c^*$ ~meson, one has transverse polarized $J/\psi$ ~meson
($\alpha=0.40$ for $B_c^*\to J/\psi\rho$ and $\alpha\approx 0.45$ for
$B_c^*\to J/\psi\pi$). On the contrary, in decays of $B_c$ ~meson, one has
longitudinal polarized $J/\psi$ ~meson
($\alpha=-0.85$ for $B_c\to J/\psi\rho$ and
$\alpha\approx -1.0$ for $B_c\to J/\psi \pi$).

The $J/\psi$ ~meson retains initial $B_c^*$ ~meson polarization in
the decay $B_c^*\to J/\psi \pi$ complitely and almost complitely (97 \%)
in the decay $B_c\to J/\psi \rho$.

The authors thank V.V. Kiselev for valuable information on the new results
concerning calculations of the decay widths of
$B_c$ ~meson.
This work is supported by the Russian Foundation for Basic Researches,
Grant
98-15-96040, and the Russian Ministry of Education, Grant
98-0-6.2-53.

\newpage

\subsection*{Figure captions}
Fig. 1: The Feynman diagrams for decays $B_c^*(B_c)\to J/\psi \rho (\pi)$.

\begin{figure}[p]
\psfig{figure=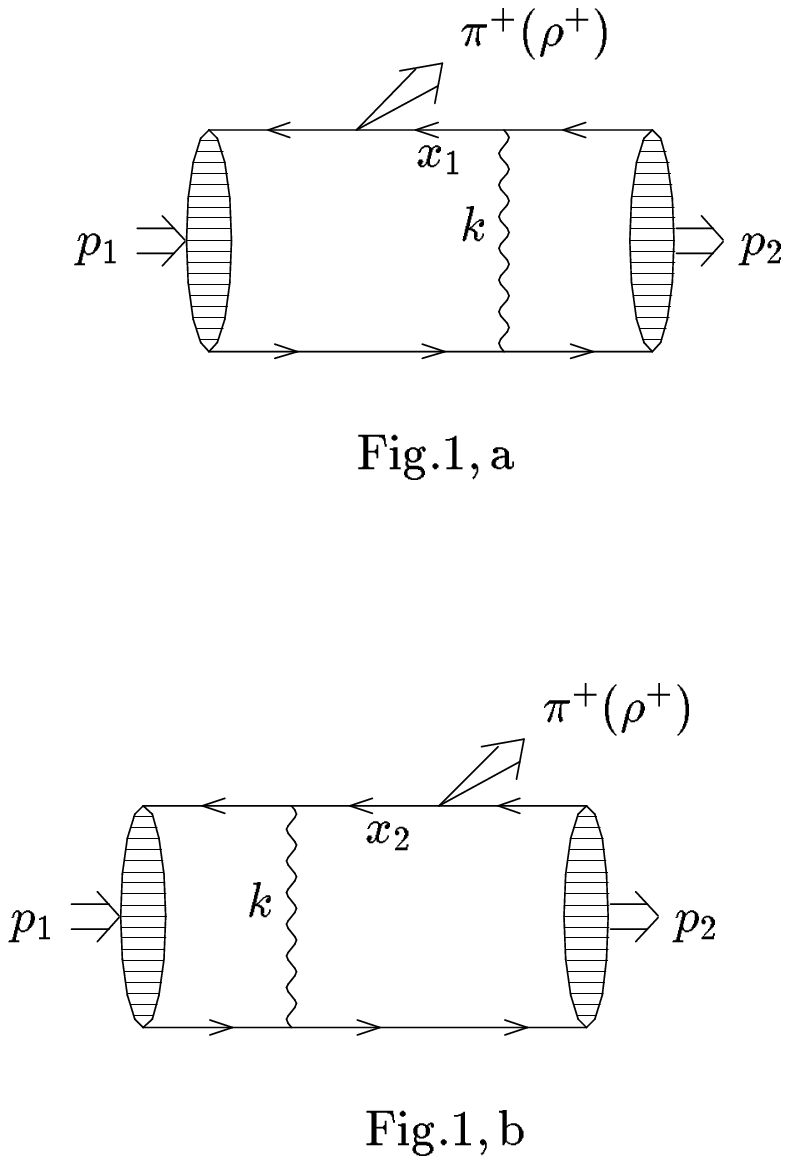,height=10cm,width=10cm,%
        bbllx=2cm,bblly=10cm,bburx=12cm,bbury=20cm}%
\end{figure}

\end{document}